\documentclass[10pt]{iopart}
\usepackage{iopams}

\usepackage{gensymb,bbm,wasysym}

\usepackage{graphicx} 
\usepackage{bm}       
\usepackage{tensind}  
\usepackage{color}

\def\ellf{\mathcal{F}}   
\def\jsn{\mbox{sn}}

\begin{document}
\tensordelimiter{?}

\title{Interactive visualization of a thin disc around a Schwarzschild black hole}

\author{Thomas M{\"u}ller}
\address{
  Visualisierungsinstitut der Universit{\"a}t Stuttgart (VISUS),\\
  Allmandring 19, 70569 Stuttgart, Germany
}
\ead{Thomas.Mueller@vis.uni-stuttgart.de}

\author{J{\"o}rg Frauendiener}
\address{
  Department of Mathematics \& Statistics, University of Otago,\\
  P.O. Box 56, Dunedin 9010, New Zealand
}
\ead{joergf@maths.otago.ac.nz}

\begin{abstract}
  In the first course of general relativity, the Schwarzschild spacetime is the most discussed analytic solution to Einstein's field equations. Unfortunately, there is rarely enough time to study the optical consequences of the bending of light for some advanced examples. In this paper, we present how the visual appearance of a thin disc around a Schwarzschild black hole can be determined interactively by means of an analytic solution to the geodesic equation processed on current high performance graphical processing units. This approach can, in principle, be customized for any other thin disc in a spacetime with geodesics given in closed form. The interactive visualization discussed here can be used either in a first course of general relativity for demonstration purposes only or as a thesis for an enthusiastic student in an advanced course with some basic knowledge of OpenGL and a programming language.
\end{abstract}


\submitto{\EJP}

\section{Introduction}\label{sec:intro}
The visual appearance of a thin accretion disc around a black hole is usually determined by direct integration of the geodesic equation from the observer to the disc. Unfortunately, this straightforward approach is numerically expensive and an interactive exploration of visual effects is cumbersome. 
However, as light rays are independent of each other, the integration can be easily parallelized using many core systems, compute clusters, or one of today's graphical processing units (GPUs). If, additionally, there is an analytic solution to the geodesic equation, an interactive visualization becomes feasible. 

In this paper we take advantage of the analytic solution for light rays in the Schwarzschild spacetime together with the highly parallel structure of a GPU for the interactive visualization of the optical appearance of a thin accretion disc.

The first visualization of a thin disc around a Schwarzschild black hole was done already by Luminet~\cite{luminet1979} in 1979. He considered the thin disc being composed of idealized particles moving on timelike circular geodesics around the black hole and isotropically emitting light based on the radiation profile by Page and Thorne (1974)~\cite{pageThorne}. Here, the mass of the thin disc does not influence the spacetime curvature and there is no self-illumination. 
The bolometric photograph of the disc is dominated by the Doppler-shift and by the resulting observed flux that transforms with the inverse $4$th power of the Doppler-factor.
Fukue and Yokoyama (1988)\cite{fukue1988} extended the visualization of Luminet to multiple wavelengths and compared an X-ray or optical photograph with a bolometric photograph. 
They also discussed light curves when an object temporarily eclipses the disc. 
The visualization by Armitage and Reynolds (2003)~\cite{armitage2003} shows the temporal variability of accretion discs around Schwarzschild black holes using magnetohydrodynamic simulations.

If there is an analytic solution to the geodesic equation of the spacetime under consideration, the intersection of a light ray with the disc can be determined immediately.
This so called emitter-observer problem, which is a boundary value problem for the connecting geodesics between the emitting point of the disc and the observer, was already discussed in the context of Kerr geodesics by Viergutz (1993)~\cite{viergutz1993} or Beckwith and Done (2005)~\cite{beckwith2005}. 
Details about the analytic solution of geodesics in the Schwarzschild spacetime can be found in Chandrasekhar~\cite{chandrasekhar} or \v{C}ade\v{z} and Kosti\'c (2005)~\cite{cadez2005}. A related situation to the thin-disc scenery, namely the analytic observation of a single timelike circular geodesic around a Schwarzschild black hole, was discussed by M{\"u}ller~\cite{mueller2009grg}.

The structure of this paper is as follows. In Sec.~\ref{sec:thindisk} we present the {\em black hole -- thin disc} scenery and show how the intersection calculation can be reduced to the two dimensional situation. In Sec.~\ref{sec:analytic}, we briefly outline the analytic solution to the null geodesic equation. In Sec.~\ref{sec:intVis}, we describe the GPU implementation. We finish with a brief discussion.

Our application is based on the Open Graphics Library (OpenGL) and the OpenGL Shading Language (GLSL)~\cite{opengl}. The source code of our cross-platform implementation using the QT~\cite{qt} framework is freely available for Linux and Windows systems from
{http://www.vis.uni-stuttgart.de/relativity}.
There is also a version that can be started directly in a web browser if WebGL~\cite{webgl} is supported. 

%
\section{Black hole -- thin disc scenery}\label{sec:thindisk}
The metric of the Schwarzschild black hole spacetime in spherical coordinates $x^{\mu}=(t,r,\vartheta,\varphi)$ is defined by the line element
\begin{equation}
 ds^2 = -\left(1-\frac{r_s}{r}\right)c^2dt^2 + \frac{dr^2}{1-r_s/r} + r^2d\Omega^2,
\end{equation}
where $d\Omega^2=d\vartheta^2 + \sin^2\vartheta\,d\varphi^2$ is the spherical surface element, $r_s=2GM/c^2$ is the Schwarzschild radius, $G$ is Newton's gravitational constant, $M$ is the mass of the black hole, and $c$ is the speed of light, see e.g. Rindler~\cite{rindler}. 

\begin{figure}[ht]
 \includegraphics[width=0.9\textwidth]{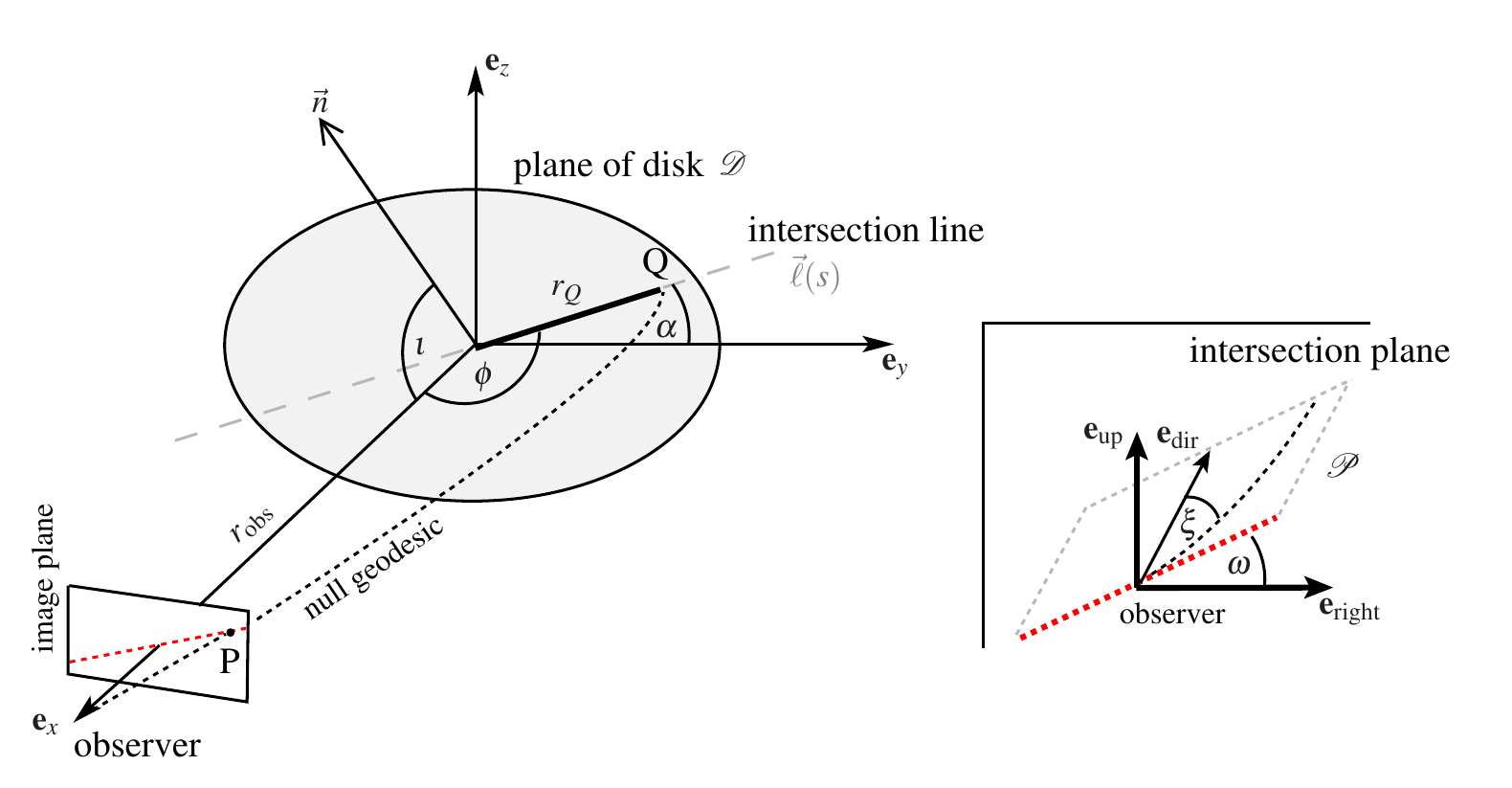}
 \caption{For each pixel $P$ of the virtual image plane of an observer located at $r=r_{\mbox{\tiny \tiny obs}}$, the intersection $Q$ of the corresponding null geodesic and the thin disc with inclination $\iota$ has to be determined. With respect to the observer's camera system, the intersection plane is tilted by the angle $\omega$ and the geodesic has initial angle $\xi$.}
 \label{fig:thindisksit}
\end{figure}

The geometric outline of the {\em black hole -- thin disc} scenery is shown in Fig.~\ref{fig:thindisksit}, where we use pseudo-Cartesian coordinates as global reference system $\mathcal{G}=\left\{\mathbf{e}_x,\mathbf{e}_y,\mathbf{e}_z\right\}$. The transformation between the spherical Schwarzschild and pseudo-Cartesian coordinates is as usual: $x=r\sin\vartheta\cos\varphi$, $y=r\sin\vartheta\sin\varphi$, $z=r\cos\vartheta$.

In Fig.~\ref{fig:thindisksit}, the Schwarzschild black hole is located at the origin. The disc with inner and outer radii $R_{\mbox{\tiny in}}$ and $R_{\mbox{\tiny out}}$, respectively, is tilted towards the observer with normal lying in the $(\mathbf{e}_x,\mathbf{e}_z)$ plane, $\vec{n}=(\cos\iota,0,\sin\iota)_{\mathcal{G}}^T$. 
The observer's position is on the $\mathbf{e}_x$ axis a coordinate distance $r_{\mbox{\tiny obs}}$ from the black hole apart.

%
\subsection{Camera system and geometry reduction} 
\begin{figure}[ht]
 \includegraphics[scale=0.6]{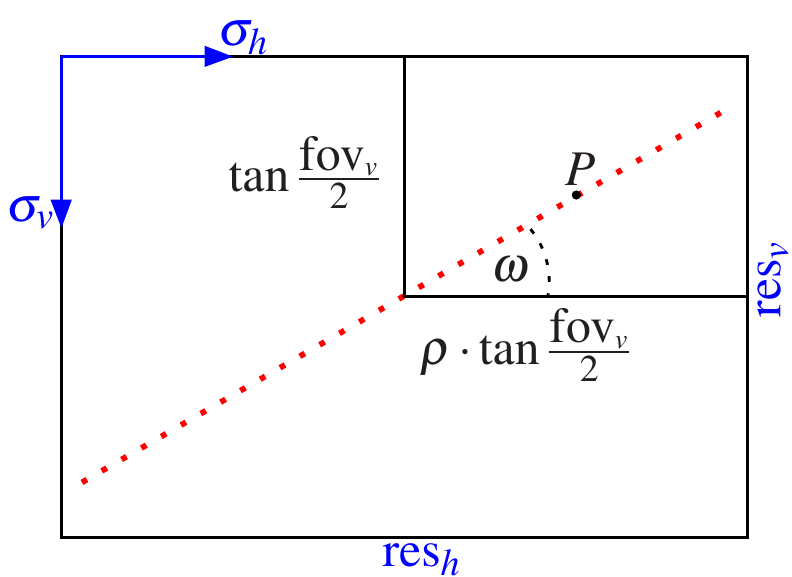}
 \caption{The camera's virtual image plane is defined by the image resolution $(\mbox{res}_h,\mbox{res}_v)$ given in pixels and the vertical field of view $(\mbox{fov}_v)$ given in degrees. A pixel $P$ is defined by its pixel coordinates $p_h\in [0,\mbox{res}_h)$ and $p_v\in [0,\mbox{res}_v)$. The origin for the pixel coordinates is the upper left corner.}
 \label{fig:pixToCoord}
\end{figure}

The reference system $\mathcal{C}$ of the observer's camera system, 
\begin{equation}
 \mathcal{C} = \left\{\mathbf{e}_{\mbox{\tiny dir}}=-\mathbf{e}_{(r)},\mathbf{e}_{\mbox{\tiny right}}=\mathbf{e}_{(\varphi)}, \mathbf{e}_{\mbox{\tiny up}}=-\mathbf{e}_{(\vartheta)}\right\},
\end{equation}
is given with respect to the local reference frame
\begin{eqnarray}
  \hspace*{-2cm}
  \mathbf{e}_{(t)} = \frac{1}{c\,\sqrt{1-r_s/r}}\partial_t, \quad
  \mathbf{e}_{(r)} = \sqrt{1-\frac{r_s}{r}}\partial_r,\quad 
  \mathbf{e}_{(\vartheta)} = \frac{1}{r}\partial_{\vartheta}, \quad \mathbf{e}_{(\varphi)} = \frac{1}{r\sin\vartheta}\partial_{\varphi}
  \label{eq:LT}
\end{eqnarray}
at his position. Here, the observer's four-velocity $\mathbf{u}=\mathbf{e}_{(t)}$, which means that he is static with respect to the asymptotic infinity. The viewing direction $\mathbf{e}_{\mbox{\tiny dir}}$ points to the black hole.
The camera's virtual image plane is defined by the vertical field of view $(\mbox{fov}_v)$ in degrees and the image resolution $(\mbox{res}_h,\mbox{res}_v)$ in pixels with aspect ratio $(\rho=\mbox{res}_h/\mbox{res}_v)$, see Fig.~\ref{fig:pixToCoord}.

\begin{figure}[ht]
 \includegraphics[scale=0.46]{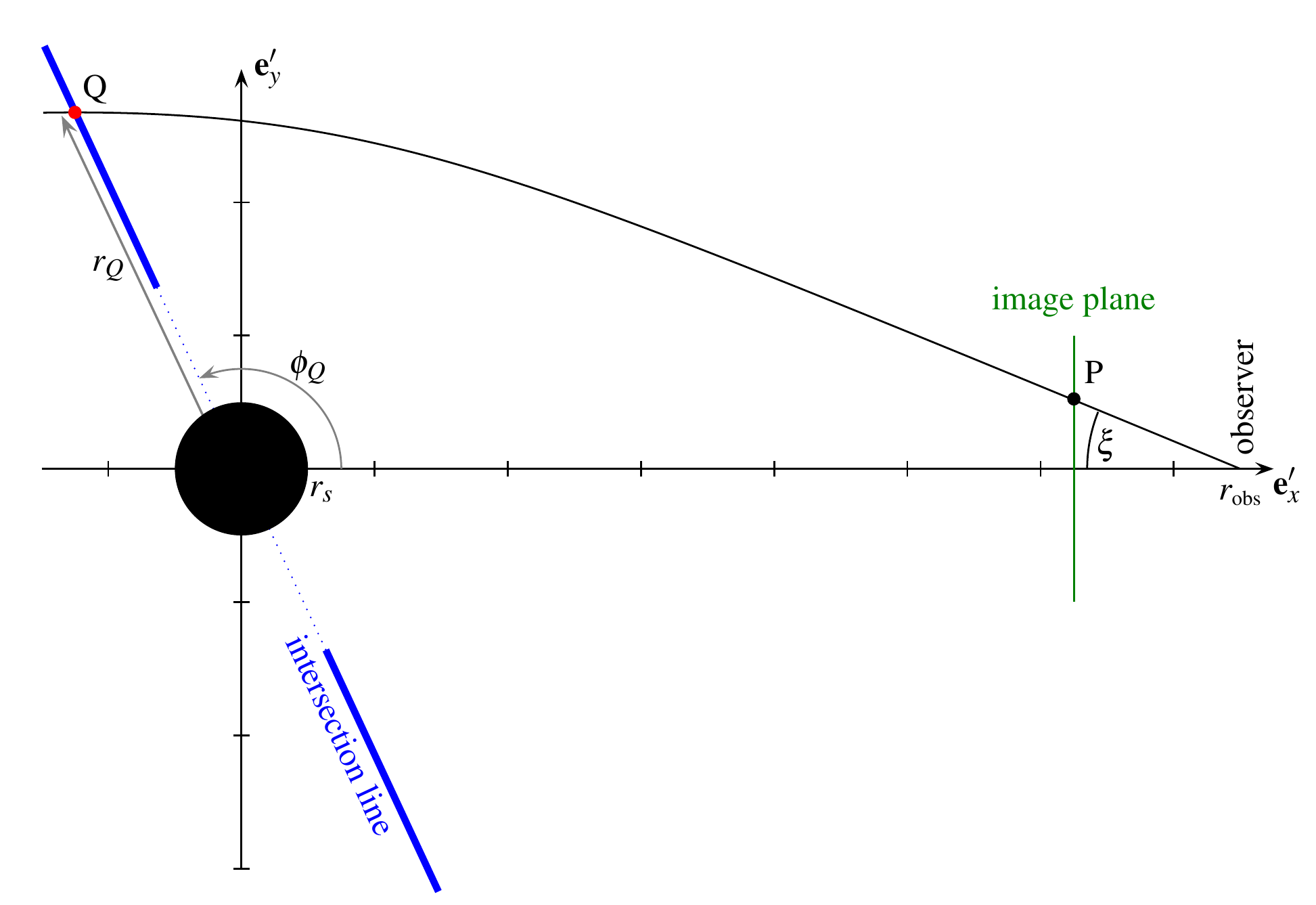}
 \caption{A null geodesic with initial angle $\xi$ in the plane $\mathcal{P}$ intersects the disc plane $\mathcal{D}$ in the point $Q$. The observer is located at $r_{\mbox{\tiny obs}}=15r_s$ and the disc radii read $R_{\mbox{\tiny in}}=3r_s$ and $R_{\mbox{\tiny out}}=7r_s$. The intersection line follows from $\mathcal{P}\cap\mathcal{D}$.}
 \label{fig:geodInSecPlane}
\end{figure}

For each pixel $P=(p_h,p_v)$ of the camera's virtual image plane, a past-directed null geodesic with initial four-direction $\mathbf{k}=(-1,\vec{k}/\|\vec{k}\|)_{\mathcal{C}}=(-1,k^d,k^r,k^u)_{\mathcal{C}}$ with respect to the observer's camera system and
\begin{equation}
  \vec{k} = \left[1,\rho\left(2\frac{p_h}{\mbox{res}_h}-1\right)\tan\frac{\mbox{fov}_v}{2}, \left(1-2\frac{p_v}{\mbox{res}_v}\right)\tan\frac{\mbox{fov}_v}{2}\right]^T,
\end{equation}
has to be tested for intersection with the disc. The minus sign in the $0$-component of $\mathbf{k}$ indicates that the null geodesic is past-directed, and the components $k^d$, $k^r$, $k^u$ are normalized.
Because in the spherically symmetric Schwarzschild spacetime the path of a null geodesic stays in a fixed two-dimensional plane, we can restrict the intersection calculation to the two-dimensional situation shown in Fig.~\ref{fig:geodInSecPlane}.

Here, the plane $\mathcal{P}$ of the null geodesic in pseudo-Cartesian coordinates is tilted by the angle $\omega$ around the $\mathbf{e}_x$ axis, see also Fig.~\ref{fig:pixToCoord}, where $\tan\omega=k^u/k^r$, and $\mathbf{e}'_x=\mathbf{e}_x$, $\mathbf{e}'_y=\cos\omega\,\mathbf{e}_y+\sin\omega\,\mathbf{e}_z$, $\mathbf{e}'_z=-\sin\omega\,\mathbf{e}_y+\cos\omega\,\mathbf{e}_z$.
The intersection between $\mathcal{P}$ and the plane of the disc, $\mathcal{D}$, is given by the straight line
\begin{equation}
 \vec{\ell}(s) = s\left(-k^u\tan\iota,k^r,k^u\right)_{\mathcal{G}}^T.
\end{equation}
The angle $\phi$ between the $x$-axis and $\vec{\ell}(s)$ reads
\begin{equation}
 \cos\phi = -\frac{k^u\tan\iota}{\sqrt{(k^r)^2+(1+\tan^2\iota)(k^u)^2}}.
\end{equation}
Hence, an intersection point $Q$ has coordinates
$(r_Q=\|\vec{\ell}(s_Q)\|,\phi_Q=\phi)$ with respect to the plane $\mathcal{P}$. 
While $\phi$ is fixed by the intersection line, we have to determine the coordinate distance $r_Q$ of $Q$ to the black hole and test whether it lies in the range $[R_{\mbox{\tiny in}},R_{\mbox{\tiny out}}]$. 
For that, we need the initial direction $\xi$ of the null geodesic within the intersection plane,
\begin{equation}
 \cos\xi = \frac{\langle\vec{k},\vec{r}\rangle_{\mathcal{C}}}{\|\vec{k}\|\cdot\|\vec{r}\|} = \frac{1}{\sqrt{1+(k^r)^2+(k^u)^2}},
 \label{eq:initXi}
\end{equation}
where $\vec{r}=(1,0,0)_{\mathcal{C}}^T$ with respect to $\mathcal{C}$. The distance $r_Q$ then follows from the analytic geodesic as discussed in Sec.~\ref{sec:analytic}.

As the gravitational frequency-shift between particle and observer depends only on the relative distances $r_Q$ and $r_{\mbox{\tiny obs}}$ to the black hole, we can go ahead and determine $1+z_{\mbox{\tiny grav}}=\sqrt{(1-r_s/r_{\mbox{\tiny obs}})/(1-r_s/r_Q)}$ (see e.g. Wald~\cite{wald}).

%
\subsection{Disc particles}
Like Luminet~\cite{luminet1979}, we consider the disc as being composed of idealized particles moving on timelike circular geodesics and radiating 
isotropically. To determine the Doppler-shift, we need the angle between the particle's direction of motion and the outgoing light ray that reaches the observer. The outgoing light direction will be discussed in the next section.

At the intersection point the non-normalized direction $\vec{m}=\vec{n}\times\vec{\ell}$ of a particle on a timelike circular geodesic around the black hole reads
\begin{equation}
 \vec{m} = \left(-k^r\sin\iota,-k^u(\sin\iota\tan\iota+\cos\iota),k^r\cos\iota\right)_{\mathcal{G}}^T,
\end{equation}
which we have to normalize and project into the rotated system $\left\{\mathbf{e}'_x,\mathbf{e}'_y,\mathbf{e}'_z\right\}$.
Depending on the distance $r_Q$ to the black hole, the particle will have four-velocity
\begin{equation}
 \mathbf{u} = c\gamma\left[\frac{1}{c\sqrt{1-r_s/r}}\partial_t+\beta\left(m'^r\sqrt{1-\frac{r_s}{r}}\partial_r+m'^{\vartheta}\partial_{\vartheta}+m'^{\varphi}\partial_{\varphi}\right)\right]
 \label{eq:fourVel}
\end{equation}
with $\gamma=1/\sqrt{1-\beta^2}$ and local three-velocity (see e.g. M{\"uller} and Boblest (2011)~\cite{mueller2011a})
\begin{equation}
 \beta = \frac{1}{\sqrt{2(r/r_s-1)}}.
\end{equation}
In Eq.~(\ref{eq:fourVel}), $(m'^r,m'^{\vartheta},m'^{\varphi})$ is the spherical coordinate representation of the normalized direction $\vec{m}'$ in the primed system, where we had to replace $m'^r$ by $m'^r\sqrt{1-r_s/r}$ to comprise the Schwarzschild coordinates instead of the spherical Minkowski coordinates.
Please note that the particle's trajectory appears as an ellipse and thus can have a radial component in the primed coordinates even though it is on a circular orbit in the global reference system. 

%
\subsection{Radiation of the disc}
The bolometric flux of radiation from a disc around a Schwarzschild black hole is given by
\begin{eqnarray}
  \hspace*{-2cm} F = \frac{F_0}{(R-3/2)R^{5/2}}\Bigg[\sqrt{R}-\sqrt{3} +\frac{\sqrt{3/2}}{2}\ln\left(\frac{\sqrt{R}+\sqrt{3/2}}{\sqrt{R}-\sqrt{3/2}}\frac{\sqrt{3}-\sqrt{3/2}}{\sqrt{3}+\sqrt{3/2}}\right)\Bigg]
\end{eqnarray}
with $F_0=3GM\dot{M}/(8\pi r_s^3)$, accretion rate $\dot{M}$, and $R=r/r_s$. The maximum $F_{\mbox{\tiny max}}/F_0\approx 9.167\cdot 10^{-4}$ is located at $R_{\mbox{\tiny max}}\approx 4.776$.
The observed bolometric flux $F_{\mbox{\tiny obs}}$ is related to the flux $F_{\mbox{\tiny src}}$ emitted by the disc via the transformation $F_{\mbox{\tiny obs}}=F_{\mbox{\tiny src}}/(1+z)^4$. 
Instead of the bolometric flux, we can also consider the observed temperature $T_{\mbox{\tiny obs}}=(F_{\mbox{\tiny obs}}/\sigma)^{1/4}$ which follows from the Stefan-Boltzmann law with constant $\sigma$.

A detailed discussion of this topic is out of the scope of this article. The interested reader might consult the underlying papers by Page and Thorne~\cite{pageThorne}, Luminet~\cite{luminet1979}, or Fukue and Yokoyama~\cite{fukue1988}. 

%
\section{Analytic null geodesic}\label{sec:analytic}
The intersection point $Q$ between the null geodesic and the disc can be determined using the Euler-Lagrange formalism (see e.g. Rindler\cite{rindler}) with the Lagrangian
\begin{equation}
 \mathcal{L} = -\left(1-\frac{r_s}{r}\right)c^2\dot{t}^2+\frac{\dot{r}^2}{1-r_s/r}+r^2\dot{\phi}^2
\end{equation}
for geodesics in the $\vartheta=\pi/2$ hypersurface of the Schwarzschild spacetime. Here, the dot represents the derivative with respect to the affine parameter $\lambda$, i.e. $\dot{r}=dr/d\lambda$.
From the Euler-Lagrange equations we obtain $\dot{r}^2+(1-r_s/r)h^2/r^2=k^2/c^2$ with the two constants of motion $k=(1-r_s/r)c^2\dot{t}$ and $h=r^2\dot{\phi}$.
For a null geodesic with initial angle $\xi$ with respect to the observer's reference frame, Eq.~(\ref{eq:LT}), these constants read $k = \pm c\sqrt{1-r_s/r_{\mbox{\tiny obs}}}$ and $h=r_{\mbox{\tiny obs}}\sin\xi$.
The coordinate transformation $\zeta=r_s/r$ of the $\dot{r}$ Euler-Lagrange equation together with $\mathcal{L}=0$ for null geodesics yields the orbital equation
\begin{equation}
  \frac{\dot{\zeta}^2}{\dot{\phi}^2}=\left(\frac{d\zeta}{d\phi}\right)^2=a^2-(1-\zeta)\zeta^2\quad\mbox{with}\quad a^2=\zeta_{\mbox{\tiny obs}}^2\frac{1-\zeta_{\mbox{\tiny obs}}}{\sin^2\xi}.
 \label{eq:orbitEq}
\end{equation}
The M\"obius transformation $\zeta=(\zeta_1\tau^2-\zeta_2)/(\tau^2-1)$ brings Eq.~(\ref{eq:orbitEq}) into the standard form of an elliptic integral of the first kind, 
\begin{equation}
  \phi = \pm\frac{2}{\sqrt{\zeta_2-\zeta_3}}\int\frac{d\tau}{\sqrt{(1-\tau^2)(1-m^2\tau^2)}},
\end{equation}
see e.g. Lawden~\cite{lawden}. The corresponding primitive reads
\begin{equation}
 \phi=\pm\frac{2}{\sqrt{\zeta_2-\zeta_3}}\ellf\left(\sqrt{\frac{\zeta-\zeta_2}{\zeta-\zeta_1}},m\right),
 \label{eq:ellip}
\end{equation}
with module $m^2=(\zeta_1-\zeta_3)/(\zeta_2-\zeta_3)$ and the three roots $\zeta_j$ of the cubic equation $a^2-(1-\zeta)\zeta^2=0$ of Eq.~(\ref{eq:orbitEq}) which can be solved using Cardano's method. 
Depending on the algebraic sign of the discriminant $D=a^2(a^2/4-1/27)$ of the cubic equation, these roots read
\begin{eqnarray}
  \hspace*{-2cm}\zeta_{1,2}=\rho\left(\cos\frac{\psi}{3}\pm\sqrt{3}\sin\frac{\psi}{3}\right)+\frac{1}{3},\quad \zeta_3=-2\rho\cos\frac{\psi}{3}+\frac{1}{3},\quad D\leq 0,\\
  \hspace*{-2cm}\zeta_{1,2}=\rho\left(\cosh\frac{\psi}{3}\pm i\sqrt{3}\sinh\frac{\psi}{3}\right)+\frac{1}{3},\quad \zeta_3=-2\rho\cosh\frac{\psi}{3}+\frac{1}{3},\quad D>0
\end{eqnarray}
with the auxiliary variables $q=a^2/2-1/27$, $\rho=\mbox{sign}(q)/3$ and $\cos\psi=q/\rho^3$ for $D\leq 0$ and $\cosh\psi=q/\rho^3$ for $D>0$.
The order of the roots can be changed at will.

Since we already know the angle $\phi$ from the intersection line, we need the inverse of Eq.~(\ref{eq:ellip}). 
With the Jacobi-sn function being the inverse of the Elliptic integral function $\ellf$ and $\phi_0=\phi(\zeta=\zeta_{\mbox{\tiny obs}})$ from Eq.~(\ref{eq:ellip}), we obtain
\begin{equation}
 \zeta = \frac{\zeta_2-\zeta_1\jsn^2\left(\frac{1}{2}\sqrt{\zeta_2-\zeta_3}(\phi\pm\phi_0),m\right)}{1-\jsn^2\left(\frac{1}{2}\sqrt{\zeta_2-\zeta_3}(\phi\pm\phi_0),m\right)}.
 \label{eq:orbitEqZeta}
\end{equation}
Now, we can determine $r_Q=r_s/\zeta_Q$ from (\ref{eq:orbitEqZeta}) with $\phi=\phi_Q$, $\zeta_{\mbox{\tiny obs}}=r_s/r_{\mbox{\tiny obs}}$, and $\xi$ from (\ref{eq:initXi}). For that, we have to restate Eq.~(\ref{eq:orbitEqZeta}) such that all expressions become real using transformation rules as described in Lawden~\cite{lawden}, or we have to do complex arithmetics. (Details can be found in the source code.)

To determine the Doppler-shift, $1+z_D$, due to the motion of the disc particle, we need the null direction $\mathbf{k}=k^{\mu}\partial_{\mu}$ of the outgoing light ray at $(r_Q,\phi_Q)$. In terms of spherical coordinates in the plane of the geodesic, $\mathcal{P}$, the null direction reads
\begin{eqnarray}
 \nonumber\mathbf{k} &=& \dot{t}\partial_t+\dot{r}\partial_r+\dot{\phi}\partial_{\phi}\\
   &=& \frac{k}{c^2(1-r_s/r)}\partial_t + \sqrt{\frac{k^2}{c^2}-\left(1-\frac{r_s}{r}\right)\frac{h^2}{r^2}}\partial_r+\frac{h}{r^2}\partial_{\phi}
\end{eqnarray}
with $r=r_Q$. Then, the Doppler-shift follows from the four-velocity $\mathbf{u}=u^{\mu}\partial_{\mu}$, Eq.~(\ref{eq:fourVel}), the null direction, and the Schwarzschild metric $g_{\mu\nu}$,
\begin{equation}
  1+z_D = g_{\mu\nu}k^{\mu}u^{\nu}.
\end{equation}
Together with the gravitational shift $1+z_{\mbox{\tiny grav}}$, we obtain the overall frequency shift $1+z=(1+z_{\mbox{\tiny grav}})(1+z_D)$.

%
\section{Interactive visualization}\label{sec:intVis}
The visualization of the {\em black hole -- thin disc} scenery, Fig.~\ref{fig:thindisksit}, is well suited for implementation on current programmable graphics hardware. Because light rays can be handled independently, we can take advantage of the highly parallel GPU SIMD (single instruction multiple data) architecture which results in an enormous speed-up compared to traditional serial CPU implementations. At the same time, we can combine the simulation and the visualization for an interactive exploration of the scenery.
By means of GPU programming interfaces like OpenGL, GLSL, CUDA, or OpenCL, the visualization algorithm as described in the previous sections can be implemented in nearly the same fashion as usual C code with only a minor overhead for the interface handling.
Here, we use the graphics library OpenGL together with the shading language GLSL. The graphical user interface is based on the Qt framework~\cite{qt}, but any other framework with an OpenGL/GLSL binding could also be used.

The starting point is a simple window-filling rectangle (quad) of size $\mbox{res}_h\times\mbox{res}_v$ that represents the observer's virtual image plane, see Fig.~\ref{fig:pixToCoord}. The so called {\em fragment program} on the GPU then generates a light ray for each pixel (fragment) of the virtual image plane, determines the intersection between each light ray and the disc, calculates the corresponding frequency shift, and transforms the bolometric flux or the temperature. In the last step, either the bolometric flux is mapped to a gray value, or the apparent temperature is color-mapped according to the blackbody color representation by  Wyszecki and Stiles~\cite{wyszecki,blackbody}.

\begin{figure}[ht]
 \includegraphics[scale=0.33]{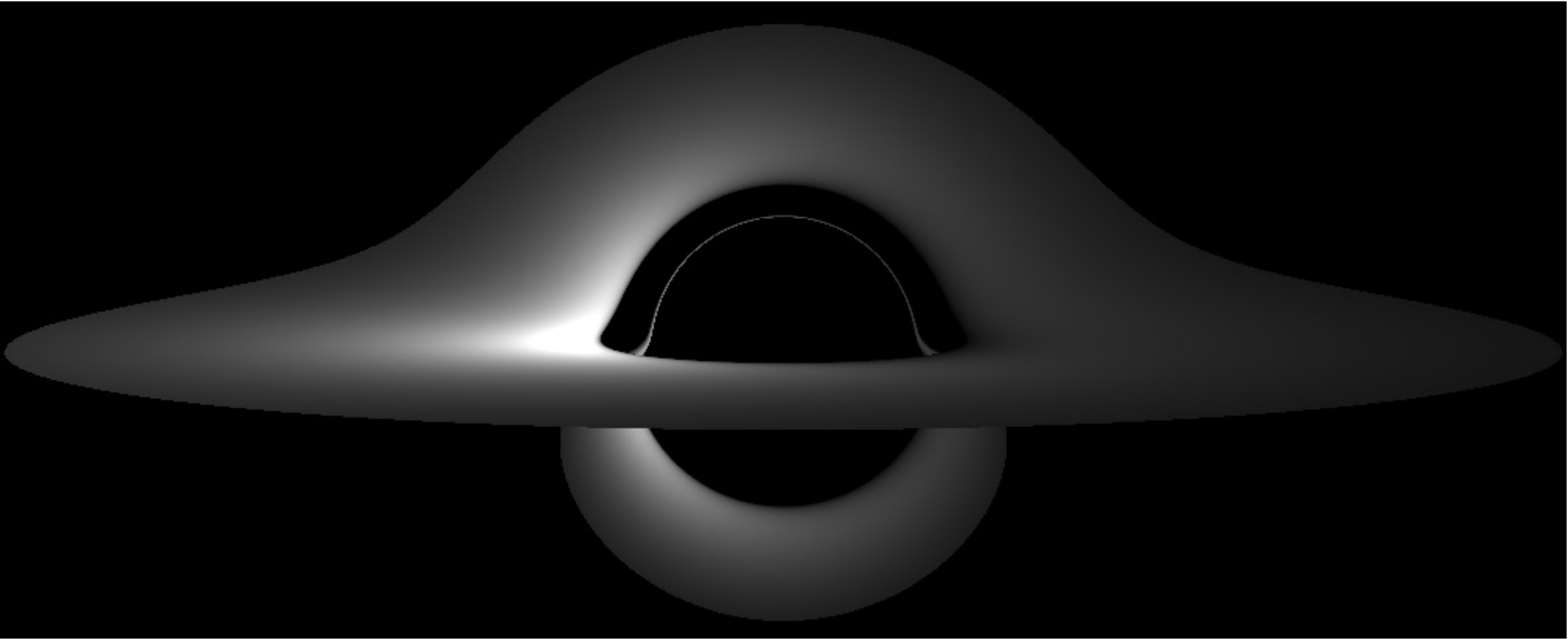}
 \caption{Visual appearance of the bolometric flux of a thin disc around a Schwarzschild black hole. The observer is located at $r_{\mbox{\tiny obs}}/r_s=120$ and has inclination $\iota=84.5\degree$ to the normal of the disc that has inner and outer radii $R_{\mbox{\tiny in}}/r_s=3$ and $R_{\mbox{\tiny out}}/r_s=15$, respectively. The camera's vertical field of view is $\mbox{fov}_v=6\degree$. For a better printout, we use a gamma correction with $\gamma=2$.}
 \label{fig:screenshot}
\end{figure}

Figure~\ref{fig:screenshot} shows the visual appearance of the bolometric flux of a narrow thin disc. Beside the primary image of the disc, there are also higher order images (lower ring and upper thin ring). The lower ring shows the rear bottom side of the disc but, because of the rotational symmetry of the Schwarzschild spacetime about the optical axis, the image is mirrored. Hence, light rays from the left part originate from the right hand side of the disc and are emitted in the direction of motion of the disc's particles. That is why the lower ring appears brighter on the left hand side.

\section{Discussion}
The main difficulty when using GPU programming is that there are still only elementary functions like square root or trigonometric functions available. More complex functions must be reimplemented by oneself. In our case, we had to implement the arithmetic-geometric mean algorithm by Abramowitz and Stegun~\cite{AbramowitzStegun64} for the Jacobi elliptic functions of Eq.~(\ref{eq:orbitEqZeta}) and some basic arithmetic operations for complex numbers.

Since our code is based on the standard graphics pipeline (OpenGL version 2.0) using only the programmable fragment unit (GLSL version 1.10), we are limited to single precision floating point calculations. However, for an observer who is not too far from the disc, $r_{\mbox{\tiny obs}}<500r_s$, this causes no significant artifacts. If double precision is essential, the presented algorithm could be easily implemented using CUDA or OpenCL.

We tested our implementation with two different graphics boards. The NVidia Geforce 8400M GT reaches about $40$ frames per second at a resolution of $600\times 500$ pixels. The more recent NVidia GeForce GTX 480, on the other hand, reaches more than $400$ frames per second at a resolution of $1000\times 1000$ pixels. 

Extensive parameter studies about relativistic scenarios that can be cut down to the emitter-observer problem like the {\em black hole -- thin disc} scenario discussed here, definitely benefit from a GPU based implementation because of the highly parallel architecture of a GPU. Even more relevant, however, is the usage of the (semi-) analytic solution to the geodesic equation that avoids the numerical integration of the geodesic equation as well as the time expensive intersection calculation.

\ack
This work was partially funded by Deutsche Forschungsgemeinschaft (DFG) as part of the 
Collaborative Research Centre SFB 716. 

\section*{References}
\bibliographystyle{unsrt}
\bibliography{lit_disk}

\end{document}